\pdfoutput=1

\documentclass[12pt]{l4dc2020} 

\usepackage{jmlrutils}
\usepackage{times}
\usepackage{graphicx}
\usepackage{amsmath}
\usepackage{float}
\usepackage{multirow}
\usepackage{siunitx}    
\usepackage{amssymb}
\usepackage{array}
\usepackage{hyperref}

\begin{document}

\author{%
 \Name{Jasmine Sekhon} \Email{js3cn@virginia.edu}\\
 \addr University of Virginia, Charlottesville, VA
 \AND
 \Name{Cody Fleming} \Email{cf5eg@virginia.edu}\\
 \addr University of Virginia, Charlottesville, VA}

\title[Vessel Intent Modeling]{A Spatially and Temporally Attentive Joint Trajectory Prediction Framework for Modeling Vessel Intent}
\date{May 2019}

\maketitle
\begin{abstract}
Ships, or vessels, often sail in and out of cluttered environments over the course of their trajectories. Safe navigation in such cluttered scenarios requires an accurate estimation of the intent of neighboring vessels and their effect on the self and vice-versa well into the future. In manned vessels, this is achieved by constant communication between people on board, nautical experience, and audio and visual signals. In this paper we propose a deep neural network based architecture to predict intent of neighboring vessels into the future for an unmanned vessel solely based on positional data. 
\end{abstract}

\begin{keywords}%
  Keywords
\end{keywords}

\section{Introduction}\label{Introduction}
Autonomous navigation is increasingly being adopted in land vehicles and airborne vehicles. The success of autonomy in other modes of travel has led to its advent in the maritime industry with the development of Autonomous Surface Vessels or ASVs. The maritime industry can potentially reap large economical and safety related benefits from these ASVs, owing to their ability to eliminate errors caused by human operators, more optimal navigation and cooperation between vessels, and trade human requirement on board with increased cargo space. However, like all other autonomous vehicles, ASVs also come with their safety and reliability concerns. These autonomous vessels, or other autonomous agents in general, are expected to negotiate safely through crowded environments, like harbors, that involve complex social interactions. %

An autonomous agent that is required to safely interact with other autonomous agents and navigate through such crowded environments must possess the ability to actively and accurately foresee the future intent of neighboring entities in order to adjust own trajectory accordingly to avoid collisions. For instance, in crowded settings such as shopping malls, train stations, etc. pedestrians use implicit rules such as respecting personal space and yielding right-of-way while moving. In the maritime domain, while safe navigation is formally governed by a set of regulations for preventing collision at sea (COLREGs), it is largely dependent on the nautical experience of people on board and their innate ability to model such implicit rules. 

The problem of predicting the future intent of a vessel based on observations of its positional data over several timesteps can be viewed as a sequence-to-sequence modeling tasks. Long Short Term Memory Networks (LSTMs), explained in greater detail in section \ref{LSTM}, are a class of deep neural networks known for their ability to model long term dependencies in a wide variety of sequence modeling tasks such as speech generation, handwriting recognition, time series prediction, and others. However, despite their success in learning and reproducing long sequences, LSTMs are not capable of modeling interactions between multiple correlated sequences such as spatially co-located autonomous agents. 

Inspired by the success of LSTMs in sequence modeling tasks and motivated by their inability to capture dependencies between correlated sequences, in this work we propose a novel temporally and spatially attentive deep learning architecture that aims to predict future intent for vessels by variably attending to observations of past spatial situations. On the surface, in our architecture, LSTM hidden states are no longer constrained to the LSTM they are associated with, and instead are also allowed to `affect' the cell states of other spatially close LSTMs. Our model is described in greater detail in section \ref{approach}. 

For an agent attempting to navigate safely in a crowded environment, the agent's \emph{domain} can be defined as the safe space surrounding the agent, the intrusion of which by any neighboring agent would cause both to have a direct impact on each other's future intent. Previous works of literature that propose to model spatial interactions in a data-driven manner assume a uniform shape for the domain and assume a uniform impact of all neighbors within the domain on the agent and vice-versa. The concept of ship domain has been crucial for safe navigation and collision-avoidance in marine transportation.  The shape of this domain has been long debated for several years. \cite{fuji_tanaka_1971} first proposed an elliptical model for ship domain for safe navigation in narrow channels. Ever since then, various shapes and sizes of ship domain models have been proposed, with variations largely due to difference in encounter situations being considered, ship dimensions, etc. (\cite{coldwell_1983, goodwin_1975, pietrzykowski_uriasz_2009, PIETRZYKOWSKI200145}). A larger fraction of the literature on ship domain uses equation systems to determine geometric dimensions of a rectangular/elliptical ship domain. These equations describe the domain as a function of features such as ship dimensions and speed etc. While such methods provide a well-defined method of calculating ship domain, they do not consider nautical experience of the navigator on board for determining ship domain/maneuvers. In our work, we propose to use data-driven methods to determine ship domain in order to take into account the non-procedural knowledge that comes from nautical experience of a navigator on board. This nautical experience is embedded in the observation data that the model is trained on. In contrast to previous data-driven approaches that rely on strong assumptions to model spatial interactions, we use this inferred domain to model the impact of a vessel on another based on their distances and relative orientations. Such insights or information about the domain along with decisions, can be used for knowledge transfer to other deep learning models or non ML models applied to the same domain. 

The goal of this work is to develop a deep learning based model to predict the future intent of socially-interacting agents. This paper:
\begin{itemize}
    \item improves on the sequence modeling capabilities of a conventional LSTM by adding the ability to model relationships between interacting sequences, such as spatially co-located autonomous agents. 
    \item introduces a novel interleaved temporal and spatial attention mechanism that enables variably attending to observations of such correlations to generate predictions. 
    \item adopts a data-driven approach for inferring useful knowledge such as ship domain based on observation data, that can be later used for knowledge transfer to other models or safety-critical domains. 
\end{itemize}

\section{Background}\label{Background}

\subsection{AIS Dataset}\label{AIS}
Most larger vessels in the U.S. or international waters transmit their position and other information such as speed and heading in real time through an onboard navigation device. This data, collected by the U.S. Coast Guard, is called the Automatic Identification System (AIS) data. The information transmitted by vessels includes the ship's unique identification number (MMSI), position in terms of latitude and longitude degrees, speed in knots, course over ground, heading and other information such as vessel name, vessel dimensions and vessel type. The frequency of transmitting this data depends on the speed and type of vessel. Since positional, course and speed history of a vessel is best reflective of normal vessel behavior, such data can prove useful to determine pro-active maneuvers to prevent potential collisions. More sophisticated data sources, such as sensors like LiDARs, radars, etc. can be used in more critical scenarios.

\subsection{Long Short Term Memory Networks}\label{LSTM}
Recurrent Neural Networks, or RNNs, are a special class of deep neural networks that are able to process input data in a sequential manner. However, RNNs suffer with the problem of vanishing and exploding gradients. Briefly, as the gradient is propagated backwards through the network's recurrent connections it tends to either shrink exponentially or blow up exponentially, which inhibits the model from learning parameters and makes the model unstable, respectively. Numerous attempts were made in the 1990s to address the problem of vanishing and exploding gradients for RNNs. On the same lines, \cite{article} introduced a special variant of RNNs called LSTMs. The primary structural difference between RNNs and LSTMs is the presence of a self-connected memory cell, whose state is regulated by three multiplicative units called gates. The \emph{input gate} regulates the flow of information from the input at time $t$ into the memory cell. The \emph{output gate} regulates the flow of information out of the memory cell towards predicting the output at $t$. The \emph{forget gate} determines the information that can be discarded from the memory cell of the LSTM. Contrary to RNNs that do not keep track of any information past the immediately previous time step, the gate-regulated cell state of LSTMs allows it to remember information from a longer history. This makes LSTMs well suited to classify, process and predict longer sequential information. Consequently, LSTMs are achieving almost human-level performance in sequence generation and are being used in a wide variety of tasks such as text generation, speech recognition, language translation, time series prediction, etc. 

\subsection{Attention Networks}\label{exp}
When trying to make a certain decision, the human brain has the natural capability to suppress idle details and focus more on certain other details. Attention networks are variants of deep learning models that mimic this capability of variably attending to different details in the input to the model. They do this by learning a \emph{weighting} over inputs or internal features that governs the flow of information through the network and consequently, the decision. These attention based models have notably shown remarkable performance in solving tasks involving sequential data, such as text generation, music generation, language translation, etc., where different parts of the input influence the prediction variably. These weights can later be extracted to infer causal relationships between different features and the outcome of the model. This is especially useful in safety-critical scenarios where a model's ability to provide sufficient reason for its behavior and decision can help improve user trust. Two variants of attention networks are relevant towards our work:\\
\textit{\textbf{Temporal Attention.}} Given a sequential input data, a future sequence generation task can be viewed as an auto-encoding task where the model first learns a fixed embedding given the input data and later uses that embedding to generate the future sequence. However, by doing this, the model assumes that every timestep in the future is uniformly dependent on timesteps in the input sequence. This leads to information loss and inaccurate predictions because in reality, in a sequence generation task, different timesteps in the observed sequence variably affect different timesteps in the future. Temporal attention is able to overcome this limitation by enabling the model to learn what to `attend' to based on the input sequence and the output it has produced so far. \cite{bahdanau2014neural} introduced a novel attention mechanism and successfully applied temporal attention to jointly translate and align words. Their attention mechanism is used in a diverse set of other tasks such as, learning alignments between image objects and agent actions in a dynamic control problem (\cite{Mnih:2014:RMV:2969033.2969073}),
between speech frames and text in the speech recognition task (\cite{Chorowski2015AttentionBasedMF}), or between visual features of a picture and its text description in the image caption generation task (\cite{xu2015attend}). \cite{luong2015effective} proposed two temporal attention mechanisms, \textit{global} and \textit{local}, that differ based on the number of input timesteps being attended to at a time. A lot of research has also focused on the alignment models used for attention. Luong et. al. compared various alignment functions such as a linear layer, dot product, concatenation, etc. In our model, we use Luong's attention, described in greater detail in section \ref{approach}. \\
\textit{\textbf{Spatial Attention.}} As mentioned earlier, a conventional LSTM lacks the ability to model interactions across sequences. In our work, we attempt to overcome this limitation by modifying the conventional LSTM architecture,  allowing the hidden state associated with an LSTM to not only recursively propagate to its own cell at the next time step, but also communicate some information about its own cell to other spatially close cells. The amount of information communicated to different cells is dependent on \textit{spatial weights}, explained in greater detail in Section \ref{hw_attn}. These \textit{spatial weights} are modeled such that spatially closer neighbors have a larger influence on the self as compared to neighbors that are spatially farther away, and are also dependent on other factors such as relative bearing and relative heading. 

\section{Approach}\label{approach}
\subsection{Motivation}
Our modeling approach is motivated by several factors, enlisted below:
\begin{itemize}
    \item As discussed in section \ref{AIS}, most vessels in the maritime domain transmit their positional, speed and course information using an AIS transmitter device on board. This sequential data is reflective of vessel behavior and its high-level trajectory, and can be used to predict future intent of the vessel. LSTMs, discussed in section \ref{LSTM}, are known to perform well in sequence prediction tasks owing to their capability of modeling long short term dependencies.
    \item Navigation in the maritime domain, not unlike other domains, involves complex interactions in a social environment. An agent, autonomous or otherwise, is required to be socially attentive to its surrounding agents and take into account their expected future behavior to negotiate a safe, collision-free path for the self. As an example, human beings are capable of being variably socially attentive to surrounding pedestrians while negotiating a collision-free path through crowded environments. For instance, they pay more attention to the present and expected future behavior of closer pedestrians and pay lesser attention to pedestrians that are far and that do not pose an immediate risk of collision. Further, influence of neighbors on the self also varies as with their orientation around the self and their direction of motion. Inherently, the default architecture of an LSTM is incapable of modeling such complex social dependencies in an interacting social environment. 
    \item Safe navigation through crowded environments also requires being variably attentive to past experiences of the self. As a maritime domain example, while trying to safely maneuver around a fishing boat, a cargo ship is expected to recollect from past experiences the expected behavior of a fishing boat, distance from the self when the fishing boat is expected to initiate a collision avoiding maneuver, the radius it turns to avoid collision, etc. 
    These past experiences of the self, are not just its own positional data, but the situations it encountered and the consequent positional decisions it made. Such variable attention mechanisms can be achieved using temporal attention, previously discussed in section \ref{exp}. 
    \item When ships approach each other, they should keep a minimum area around them clear of other vessels in order to remain safe. The geometrical shape of this area, called the vessel domain, has been heavily debated for years (\cite{fuji_tanaka_1971,goodwin_1975,PIETRZYKOWSKI200145,pietrzykowski_uriasz_2009,10.1007/0-387-26325-X_37}). A lot of prior work on trajectory prediction for vessels has assumed this ship domain to be circular or elliptical. However, the shape of a ship's domain depends on several complex factors such as the length and width of the vessel of interest, the water depth, the COLREG situation, the heading of the self and the target vessels, etc. As a part of our approach, our model infers from data, an upper bound on this vessel domain, i.e., a safe space surrounding a vessel, the intrusion of which by any other vessel would have a direct impact on the future intent of the self and the target vessel because of the possibility of a collision scenario. 
    
\end{itemize}
\subsection{Model Architecture}

Our proposed model architecture is shown in figure \ref{fig:model}. Given $N$ vessels present in a given area and actively transmitting AIS data at the beginning of an observation time window t$_\text{s}$=t$_{0}$ to t$_\text{s}$=T\textsubscript{obs}, our model uses an LSTM-based autoencoder to identically model the observed sequences of the $N$ vessels. The observed sequence for a vessel $v$ is denoted by  $\mathbf{x}^{v}_{t_{0}:T_{obs}}$ and is composed of its positional information extracted from the AIS data. The features we extract from AIS data are positional data (latitude, longitude), speed over ground and heading of the vessel. 

\subsubsection{Encoding Stage}
At each timestep t$_\text{s}$ in the observed sequence spanning over time interval [t\textsubscript{0},T\textsubscript{obs}], the hidden state of every vessel $v$, denoted by $h^{v}_{t_s}$ is updated by feeding the hidden state from the previous timestep $h^{v}_{t_{s}-1}$  and the observed features at $t_s$, $\mathbf{x}^{v}_{t_s}$ to the encoder. However, the hidden state at t$_\text{s}$ is also variably influenced by the hidden states of spatially close neighbors. As mentioned earlier, a conventional LSTM cannot take this influence into consideration. To take this spatial effect into account, we incorporate a \emph{spatial attention mechanism}, explained in greater detail in Section \ref{hw_attn}. In summary, the spatial attention mechanism aggregates together variable amount of information from hidden states of spatially close neighbors. The amount of information extracted from each neighbor is computed based on a weighting mechanism, and is influenced by different factors such as distance from $v$, relative bearing with respect to $v$ and difference in heading angle from that of $v$. The spatially-weighted hidden state of $v$, $\tilde{h}^{v}_{t_{s}-1}$ is then fed into the encoder at the next time step to update the hidden state corresponding to $v$ using the conventional update mechanism:
\begin{equation}\label{eqn:lstm}
    h^{v}_{t_{s}} = \mathbf{LSTM}(\mathbf{x}^{v}_{t_{s}-1}, \tilde{h}^{v}_{t_{s}-1})
\end{equation}

\subsubsection{Decoding Stage}
Every spatially weighted hidden state, $\tilde{h}^{v}_{t_{s}}$ corresponding to every vessel $v$ is a vector representation of the \emph{spatial situation} at $t_s$. It summarises the orientation of neighbors around $v$, their distances from $v$, their headings with respect to $v$ and their resulting influence on $v$. The decoding LSTM receives a sequence of these spatially weighted hidden states for each vessel $v$ for every $t_s$ in 
the observation time window [$t_0$,$T_{obs}$]. 

Similar to the encoding stage, for every time step $t_p$ in the prediction time window from $T_{obs}+1$ to $T_{pred}$, the decoder computes the spatial influence of the \emph{future intent} of neighbors on the \emph{future intent} of the self and vice versa using the same spatial attention mechanism. This is analogous to a pedestrian altering their path if they \emph{anticipate} collision with another pedestrian at a future time step. Further, in order to predict the intent of $v$ given a sequence of observed trajectory, it is useful to compare the \emph{anticipated} situation at every timestep $t_p$ in the prediction time window, [$T_{obs+1}$, $T_{pred}$] with the history of observed situations, $\tilde{h}^{v}_{t_{s}}$. This is similar to a pedestrian using knowledge from past experiences to determine a safe future trajectory. In the maritime domain, this is similar to a cargo ship recollecting from past experiences, the safest way to maneuver around a fishing boat when the fishing boat is present at a certain distance and relative bearing from it. Therefore, to make the model better gauge the spatial influence of the future intent of neighbors on the future intent of the self and vice versa, we interleave the spatial attention mechanism with \emph{temporal attention} mechanism. The temporal attention mechanism compares the spatially weighted hidden state at a time step $t_p$ in the prediction time window to all spatially weighted hidden states in [$t_{0}$,$T_{obs}$]. This is analogous to a vessel reacting similarly to situations it has observed previously and is used to make the model aware of similarity in spatial situations, hence enabling it to learn from the encoded input and react similarly. The temporally spatially weighted hidden state at a time step is then used to compute the hidden state corresponding to $v$ at the next time step, and the predicted intent at the next time step. The temporal attention mechanism is explained in further detail in Section \ref{sw_attn}. 
\begin{figure}%
\centering
    \includegraphics[width=0.9\textwidth]{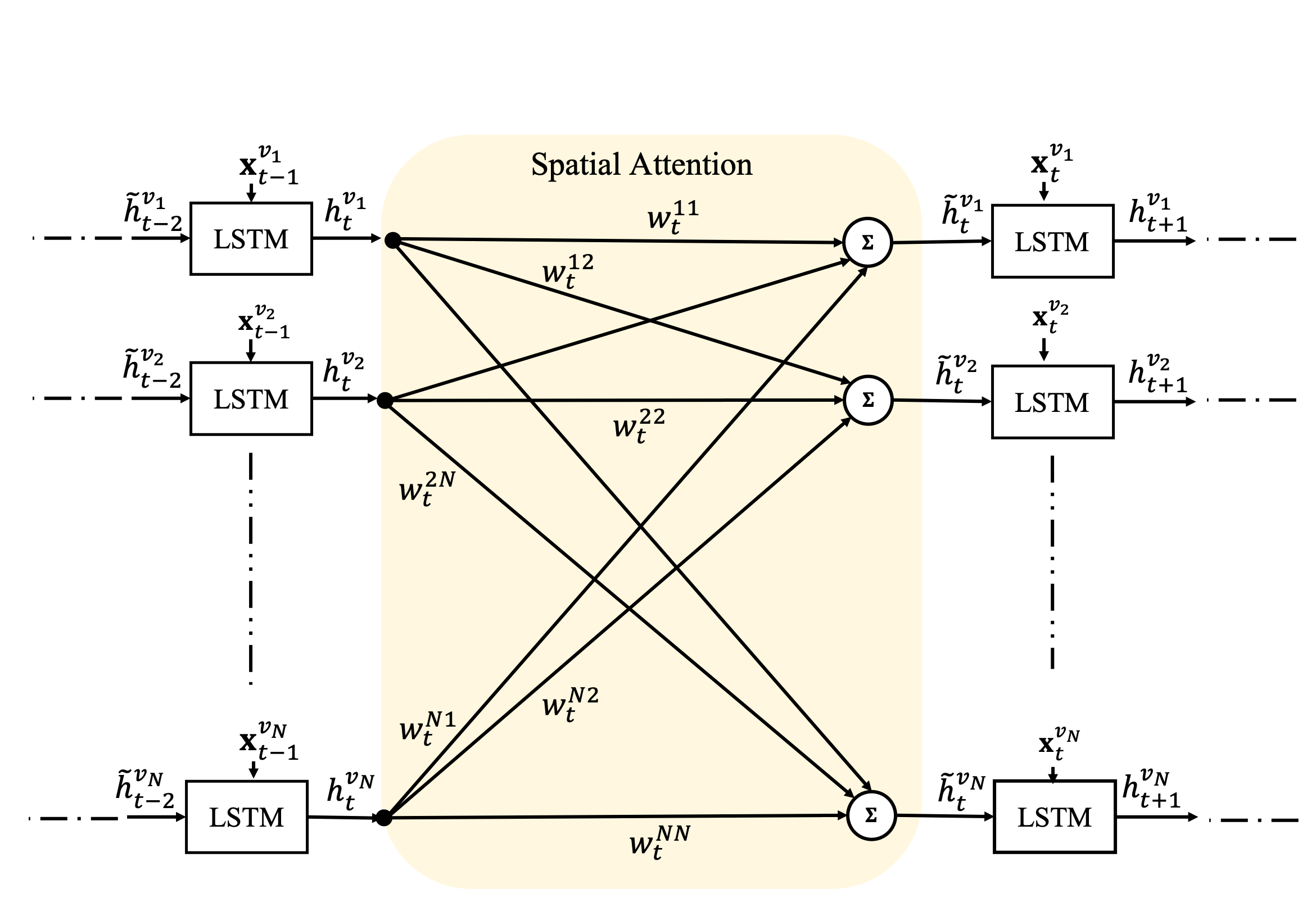}
    \caption{Spatial Attention Mechanism}\label{fig:spatial_attention}
    \end{figure}

\subsubsection{Spatial Attention}\label{hw_attn}
A socially interacting agent's intent is not only influenced variably by neighbors depending on their distance from it, it is also affected by other factors, prominent being their relative bearing from the agent and their heading angle. For instance, in the pedestrian domain, a human is most likely to be influenced by neighboring pedestrians in its line-of-sight than those behind it. In the same way, in the maritime domain, the effect of a neighbor on a vessel's intent would vary with its orientation around the vessel. On similar lines, a neighbor intending to overtake a vessel $v$ would affect the intent of $v$ differently than another neighbor approaching towards $v$ head-on, present at the same distance. To incorporate this multimodal spatial effect, we introduce a spatial attention mechanism to model the influence of neighboring vessels on the intent of the vessel of interest, and vice-versa. While data-driven approaches to vessel intent modeling are limited, several pioneering works that model human-human interaction in the pedestrian domain have introduced some forms of spatial attention (\cite{gupta2018social,Alahi_2016_CVPR,Sadeghian_2019_CVPR,Fernando2017SoftH}). However, these methods are replete with over-limiting assumptions on the (equal) number of neighbors that identically affect the intent of a pedestrian in each direction, or alternatively grid size. In contrast to these approaches, we let the model deduce the \emph{vessel domain} from the observed data. As mentioned earlier, we define the domain of a vessel as the area around the vessel, the intrusion of which by other neighboring vessels would cause a direct influence on the intent of the vessel and vice-versa. Any neighboring agent that violates this area around a vessel would be deemed as a threat to its navigational safety and would cause the vessel to initiate timely maneuvers to avoid risk of collision. Consequently, any neighboring agent beyond this area would not have any influence on the vessel and vice-versa. We denote this domain by a learn-able parameter $S$. This parameter $S$ is treated like any other trainable parameter in the model (such as weights in the encoder and decoder LSTMs) and is learned from training on observed data. Our spatial attention mechanism recognizes that at time $t$, the spatial influence of a neighboring vessel $v_2$ on a vessel, $v_1$ is dependent on three prominent factors: 
\begin{itemize}
    \item the distance of $v_2$ from $v_1$ at $t$, $d^{21}_t$
    \item the heading angle of $v_2$ with respect to $v_1$ at $t$, denoted by $\phi^{21}_t$. For example, if $v_2$ is approaching $v_1$ head-on, $\phi^{21}_t\approx180^{o}$ and if either is overtaking the other, $\phi^{21}_t\approx0^{o}$. 
    \item the relative bearing of $v_2$ with respect to the heading of $v_1$ at time $t$, denoted by $\theta^{21}_t$. The relative bearing of $v_2$ with respect to $v_1$ is the angle the line from $v_1$ to $v_2$ makes with respect to the heading angle of $v_1$. For example, if $v_2$ is approaching $v_1$ head-on, $\theta^{21}_t\approx0^{o}$ and if $v_2$ is tailgating $v_1$, $\theta^{21}_t\approx180^{o}$. 
\end{itemize}

Figure \ref{fig:spatial_attention}  shows the spatial attention mechanism. At a time $t$, the spatial influence of $v_2$ on $v_1$ is then determined by computing its spatial weight, $w^{21}_t$,
\begin{equation}
    w^{21}_t = \mathbf{ReLU}(S(\theta^{21}_t,\phi^{21}_t) - d^{21}_t)
\end{equation}

The inclusion of $S$ allows a vessel, $v_2$ farther from the corresponding value in $S$ but closer to $v_1$ to be assigned a larger weight. $\mathbf{ReLU}$ is a non-linear activation function commonly used in deep neural networks. For any input $i$, $\mathbf{ReLU}(i) = max(0,i)$. Here, this activation function ensures that if the distance of $v_2$ from $v_1$, $d^{21}_t$ is greater than the corresponding domain value $S(\theta^{21}_t,\phi^{21}_t)$, $v_2$ would have no effect on the intent of $v_1$. Both these features ensure that the trained parameter $S$ models the vessel domain. 

Using this weighting mechanism, spatial weights are calculated for every pair of vessels in the given frame. Once this is done, the \emph{spatially weighted} hidden state of $v_1$ is computed as:
\begin{equation}
    \tilde{h}^{v_1}_{t} = w^{11}_{t}h^{v_1}_{t} + w^{21}_{t}h^{v_2}_{t} + \ldots + w^{N1}_{t}h^{v_N}_{t}
\end{equation}
\begin{figure}[t]
    \centering
    \includegraphics[width=0.9\textwidth]{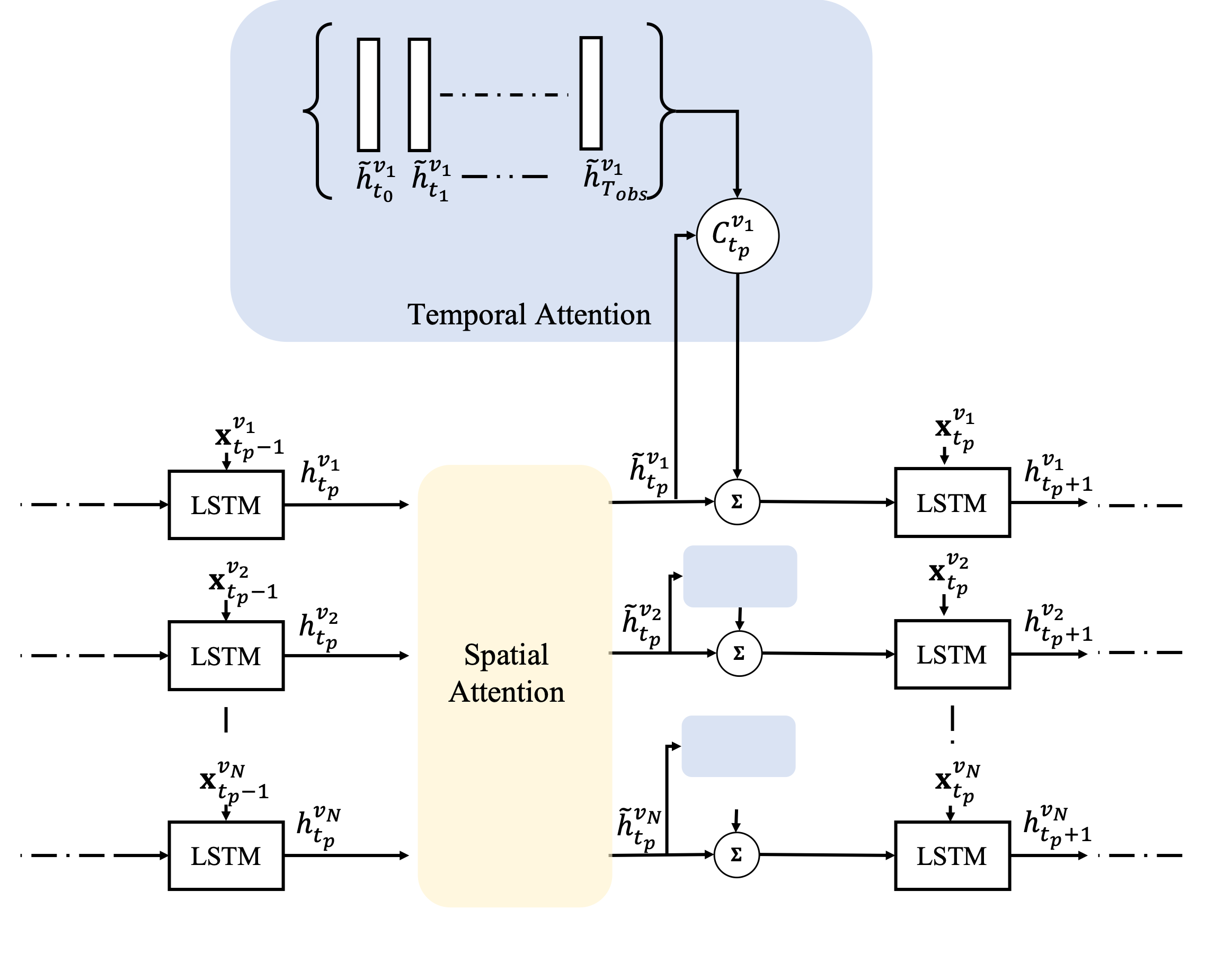}\caption{Temporal Attention Mechanism}\label{fig:temporal_attention}
    \end{figure}
where $N$ is the number of vessels co-located in the given frame. This spatially weighted hidden state is then fed to the encoder or the decoder at the next time step to update the hidden state corresponding to $v_1$, $h^{v_1}_{t+1}$.

\subsubsection{Temporal Attention}\label{sw_attn}
At every timestep $t_p$ in the prediction time window [$T_{obs+1},T_{pred}$], the decoder first uses the spatial attention mechanism to summarise the `situation' or the orientation of neighbors around $v_1$ and their influence on $v_1$ thereof. It then compares this spatially weighted hidden state $\tilde{h}^{v_1}_{t_p}$ with all $\tilde{h}^{v_1}_{t_s}$, $t_s\in$[$t_0$,$T_{obs}$], to understand from similar past experiences the best way to navigate through this situation. This is done using a temporal attention mechanism, shown in Figure \ref{fig:temporal_attention}. In our model, we specifically use the attention mechanism introduced by \cite{luong2015effective}. 

At each time step $t_p$ in the prediction sequence, the LSTM associated with $v$ computes a \emph{context vector}, $\mathbf{C}^v_{t_p}$ as the weighted sum of (spatially-weighted) hidden states from the observed time window:
\begin{equation}
    \mathbf{C}^v_{t_p} = \sum_{t_s=t_0}^{T_{obs}} = \alpha_{t_p}\tilde{h}_{t_s}^{v}
\end{equation}
\begin{figure*}[ht]
    \centering
    \includegraphics[width=\textwidth]{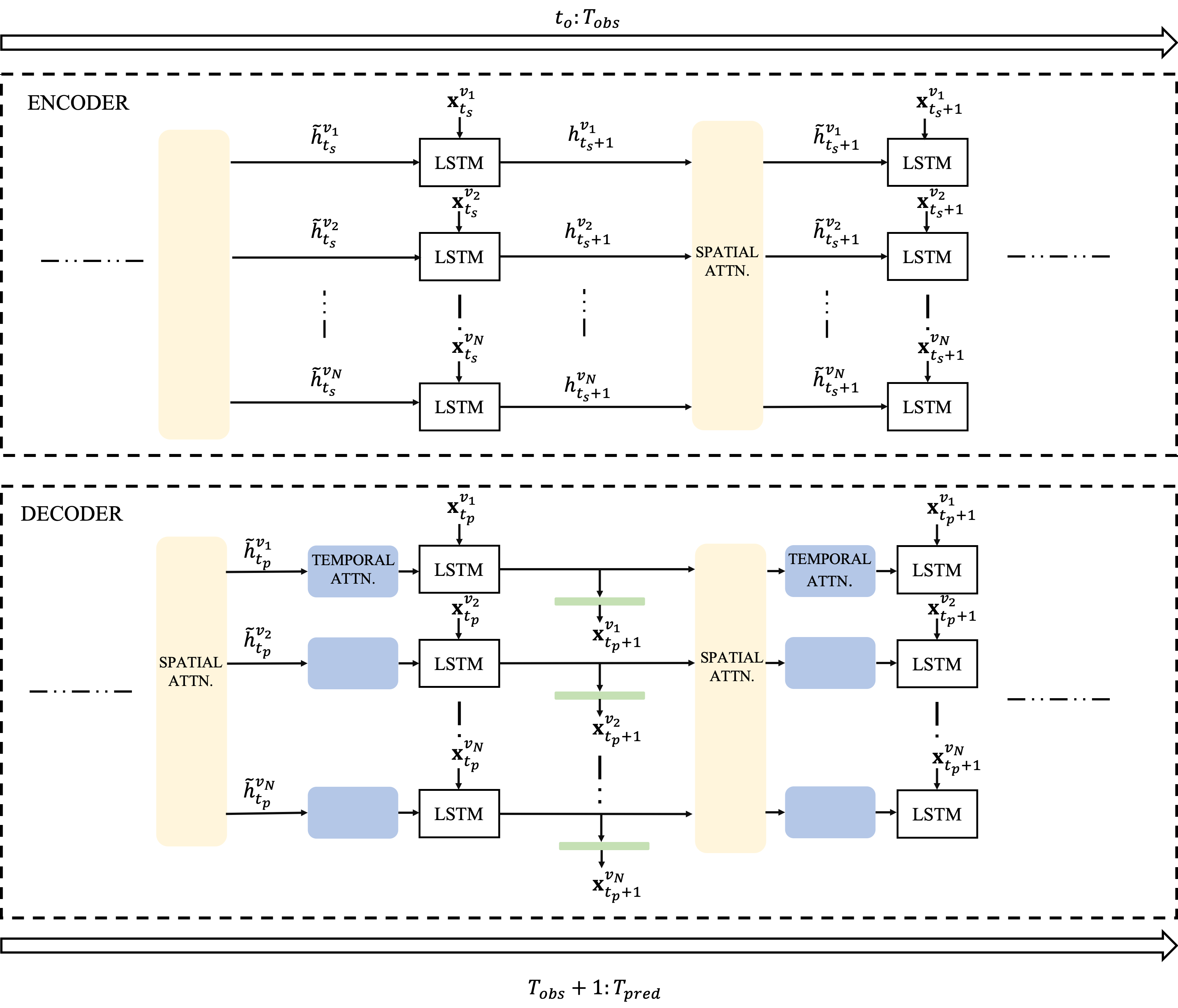}
    \caption{Spatially and Temporally Attentive LSTM-based Auto-encoder for Vessel Intent Prediction. The yellow blocks represent the spatial attention mechanism shown in Figure \ref{fig:spatial_attention}. The blue blocks represent the temporal attention mechanism shown in Figure \ref{fig:temporal_attention}. The green blocks stand for fully connected layers that convert the hidden state of the LSTM at time step $t$ into the predicted intent at $t$.}
    \label{fig:model}
\end{figure*}
The alignment vector $\alpha_{t_p}$, with length equal to the number of time steps in the observed sequence, is derived by comparing the current spatially-weighted hidden state $\tilde{h}^{p}_{t_p}$ with each spatially-weighted hidden state $\tilde{h}_{t_s}^{p}$ from the observed sequence:
\begin{equation}
    \alpha_{t_p} = \mathbf{align}(\tilde{h}^{v}_{t_s},\tilde{h}^{v}_{t_p})
\end{equation}

\begin{equation}\label{eqn:score}
    \mathbf{align}(~{h}^{v}_{t_s},~{h}^{v}_{t_p}) = \frac{\mathbf{exp}(\mathbf{score}(\tilde{h}^{v}_{t_s},\tilde{h}^{v}_{t_p})}{\sum_{s'}\mathbf{exp}(\mathbf{score}(\tilde{h}^{v}_{t_{s'}},\tilde{h}^{v}_{t_p})}
\end{equation}
where $\mathbf{score}$ is called \emph{content-based function} and is used to quantify the similarity of a source hidden state and a target hidden state. 

An observed experience or situation being identical to the current situation would cause the two spatially weighted hidden states being compared to be equal. To allow such similar observed experiences to be assigned a higher $\mathbf{score}$ in Equation \ref{eqn:score}, we use dot product to compute the $\mathbf{score}$. This is because dot product is maximum when the two hidden states being compared are `equal', which would mean that the spatial situations being summarized by the two spatially weighted hidden states being compared are identical. 
Therefore,

\begin{equation}
    \mathbf{score}(\tilde{h}^{v}_{t_{s}},\tilde{h}^{v}_{t_{p}}) = \tilde{h}^{v}_{t_{s}} \cdot \tilde{h}^{v}_{t_{p}}
\end{equation}

The soft attention context vector $\mathbf{C}^{v}_{t_p}$ is computed at every $t_p \in [T_{obs}+1 ,T_{pred}]$. At every time step, it is concatenated with the computed spatially weighted hidden state, $\tilde{h}^{v}_{t_{p}}$ and is further used to update the hidden state of the decoder according to Equation \ref{eqn:lstm} and generate an intent prediction. A fully connected linear layer is used to convert the updated hidden state into a predicted intent for $v_1$ at $t_p$. 
\begin{equation}
    \tilde{h}^{v}_{t_{p}} = \mathbf{concat}(\mathbf{C}^{v}_{t_p}, \tilde{h}^{v}_{t_{p}})
\end{equation}
\begin{equation}
    \mathbf{x}^{v}_{t_p} = \mathbf{linear}(h^{v}_{t_p})
\end{equation}
where $\mathbf{x}^{v}_{t_p}$ is the predicted position or intent at $t_p$ for $v$.

\section{Implementation}\label{Experiment}
\textbf{\textit{Dataset and Pre-processing.}} To evaluate our model, we use AIS records within U.S. coastal waters from January 2017\footnote{retrieved from \url{https://marinecadastre.gov}}. These records have been filtered to one minute intervals, with sensitive fields like ship names and call sign removed. Each AIS sample contains information pertaining to the following fields: MMSI (the unique vessel ID), BaseDateTime (time stamp), LAT (latitude), LON (longitude), SOG (speed over ground), COG (course over ground), Heading (heading angle), VesselName, IMO Number, CallSign, Vessel Type, Navigational Status, Length, Width, Draft, Cargo. While some missing and outdated information has been pre-corrected by the Authoritative Vessel Identification Service (AVIS), most of these fields are still missing in the available data. In our model, we only use MMSI, BaseDateTime, LAT, LON, SOG and Heading values. Since we are interested in being able to predict intent in crowded environments, we train and validate our model on available AIS data from 681 vessels around San Diego Harbor (UTM Zone 11) from January 2017. 
Different kinds of vessels traveling at different speeds update their AIS information at different rates. An anchored/moored vessel updates its dynamic data once every three minutes, whereas a ship with a speed of 23 knots or higher is required to update its dynamic data every 2 seconds. Since our model processes concurrent AIS information from all vessels in a certain grid space or frame, we resample the raw AIS data to one minute intervals. For vessels transmitting their AIS information at a lower rate than one minute, we use linear interpolation to fill in the missing data to up to five minutes.
All timestamps are rounded to the nearest one minute interval and duplicate timestamps are removed. Our goal is to be able to predict the intent of moving vessels that are actively engaging in spatial interactions with other moving vessels, hence having a possible effect on their intent and vice-versa. Therefore, we remove timestamps with AIS information from only moored/anchored vessels with recorded SOG values less than 1 knot and timestamps with concurrent AIS information from less than three vessels. Our filtered dataset contains AIS data from 653 vessels in and around the San Diego Harbor from January 2017.  

\textbf{\textit{Architecture Details.}} To substantiate our choice of architecture, we trained and evaluated our model in an ablative setting:
\begin{itemize}
    \item \textit{LSTM+Spatial+Temporal Attention.} This refers to our proposed model, with a spatial attention mechanism to incorporate the spatial interactions with other agents in close proximity and a temporal attention mechanism to enable the model to learn variably from its history of observed experiences. At each timestep $t_s$ in the observed time window $t_0$ to $T_{obs}$, this model uses an LSTM-based encoder to update hidden states of $N$ vessels present in the given frame using their AIS information at that time step, $x^{v}_{t_s}$, $v\in N$. At each timestep $t_s$ in the observed time window, the encoder LSTM first uses the spatial attention mechanism, as described in Section \ref{hw_attn}, to take into account neighboring vessels and their effect on the self and vice-versa. At the end of the encoding stage, each vessel $v$ is associated with weighted hidden states corresponding to each timestep $t_s$ in the observed time window, $\tilde{h}^{v}_{t_{s}}$, which is essentially the vessel's ``observed'' situation at $t_s$. 
    
    For every timestep in the decoding stage, $t_p\in$[$T_{obs+1}$,$T_{pred}$], the decoder LSTM first uses the same spatial attention mechanism to incorporate spatial influences. It then uses a temporal attention mechanism, as described in Section \ref{Background} to \emph{compare} the current ``observed situation'' for every vessel $v$ with its history of observed situations from the encoding stage. By doing so, it tries to assess the variable effect of different timesteps in the observed sequence on the intent at $t_p$. 
    \item \textit{LSTM+Spatial Attention.} This refers to our model with only the spatial attention mechanism to incorporate spatial interactions with other neighbors in close proximity. This model does not take into account temporal attention mechanism to understand the variable effect of observed situations on the predicted intent. The encoding and decoding stage for this model are essentially identical. 
    \item \textit{LSTM+Temporal Attention.} This model consists of a vanilla-LSTM with a temporal attention mechanism. This model is agnostic to spatial interactions with neighbors in close proximity while predicting intent for a certain vessel $v$. It, however, does incorporate the variable temporal effects of different timesteps in the observed time window for each vessel $v$ while predicting intent. 
    \item \textit{vanilla-LSTM.} This baseline model consists of a single-layer vanilla-LSTM that tries to model intent while being agnostic to any spatial or temporal influences. 
\end{itemize}

\textbf{\textit{Training Details.}} We iteratively trained all models for 500 epochs using the Adam optimizer with a batch size of 64 and a learning rate of 0.001. For training our model, we split all vessel trajectories into smaller windows of length equal to the observation time window, each window being one sample. The input to the models contains AIS information from an observation time window of 5 minutes (5 timesteps) and the models attempt to predict intent for all the observed vessels over the next 5 minutes. We extracted 8676 such samples from the processed AIS data, using 80\% for training, 10\% for validation and the remaining 10\% for testing the trained models. The hidden layer dimensions of the \textit{vanilla-LSTM} and the encoder and decoder LSTMs in the other three models is 6. We observed that in many cases, the recorded AIS speed and Heading values are not consistent with the recorded positional data (latitude, longitude values). Therefore, we use only two input features, i.e., latitude and longitude values. The outputs of the decoder LSTMs at each timestep in the prediction window are fed to a fully-connected layer of size 2 that infers the predicted positional intent, i.e., latitude and longitude values from the hidden state of the decoding LSTMs. For each batch in each iteration, we train the model hyper-parameters by computing loss as defined by the ADE metric in Section \ref{metrics}. Since our model predicts intent in geographic coordinates, we use equirectangular distance to compute displacement error. 

\section{Metrics}\label{metrics}
To evaluate the performance of our model on the AIS dataset, we adopt two metrics commonly used in the pedestrian domain for evaluating trajectory prediction methods(\cite{Alahi_2016_CVPR,gupta2018social,Sadeghian_2019_CVPR}). These are:
\begin{itemize}
    \item \emph{Average Displacement Error (ADE)}: It is defined as the average displacement between the predicted trajectory and ground truth trajectory over the prediction time span [$T_{obs+1},T_{pred}$] across all the vessels in the frame. 
    \item \emph{Final Displacement Error (FDE)}: It is the displacement error between the final predicted positions and ground truth positions at the end of the prediction time span, i.e. at $T_{pred}$ averaged over all the vessels in the frame. 
    \end{itemize}

\section{Evaluation}\label{eval}
We evaluate the performance of our model in different ablative settings on data from UTM Zone 11 and report ADE and FDE values (in nautical miles) in Table \ref{tab:eval_11}. Since the \textit{vanilla-LSTM} does not incorporate spatial interactions and solely uses the vessel's own observed history to predict its intent, the \textit{vanilla-LSTM} and its variant with temporal attention perform the worst. The {vanilla-LSTM + spatial attention} model is able to perform better than the models without any spatial attention mechanism because of its ability to understand the causal relationship between a vessel's neighborhood and its intent. Adding temporal attention to this model further improves performance because the model is then able to learn from past ``situations'' as observed by the self and variably attend to these while predicting intent, alongwith understanding and incorporating spatial influences. Despite the LSTM encoder and decoder being single-layer LSTMs with very small hidden dimensions of 6, our model is able to perform well because of its interleaved spatial and temporal attention mechanisms that are able to intelligently capture the complex cause-effect relationships among neighbors, their observed experiences and each vessel's individual intent. The ability to determine the distance and orientation aspect of spatial relationships via learning instead of making strong assumptions is another factor that contributes to the performance of our model. 
\begin{table}[]
    \centering
    \begin{tabular}{|c|c|c|c|c|} \hline 
    \textbf{Metric} & \textit{vanilla-LSTM} & \textit{LSTM }  & \textit{LSTM } & \textit{LSTM + Spatial + } \\
    & & \textit{+ temporal attention} & \textit{+ spatial Attention} & \textit{+ Temporal Attention} \\ \hline 
     ADE & 0.04567 & 0.04152 & 0.03912 & \textbf{0.03314} \\ 
     FDE & 0.05377 & 0.05601 & 0.04292 & \textbf{0.03840} \\ \hline 
\end{tabular}
    \caption{Quantitative Results for all models on evaluation dataset from UTM Zone 11. The ADE and FDE values are reported in nautical miles and are computed for predicted intent over 5 minutes using observed AIS information from 5 minutes.}
    \label{tab:eval_11}
\end{table}

\section{Discussion}\label{domain}
\begin{figure}%
    \centering
    \includegraphics[width=0.5\textwidth]{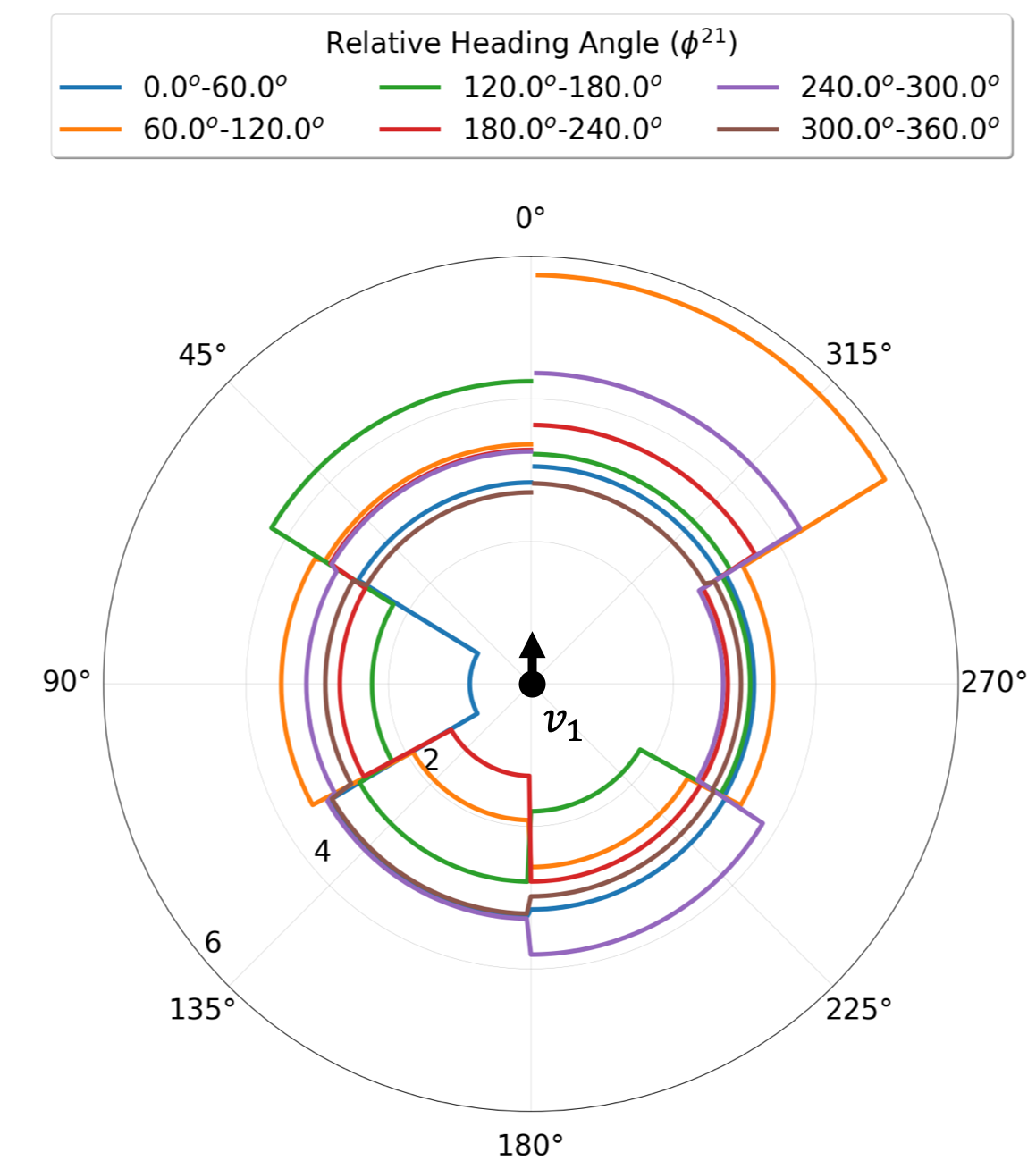} \caption{Learned Domain Parameter for various $\phi_{21}$}
    \label{domain_img}
    \end{figure}

\begin{figure}
        \centering
        \includegraphics[width=0.5\textwidth]{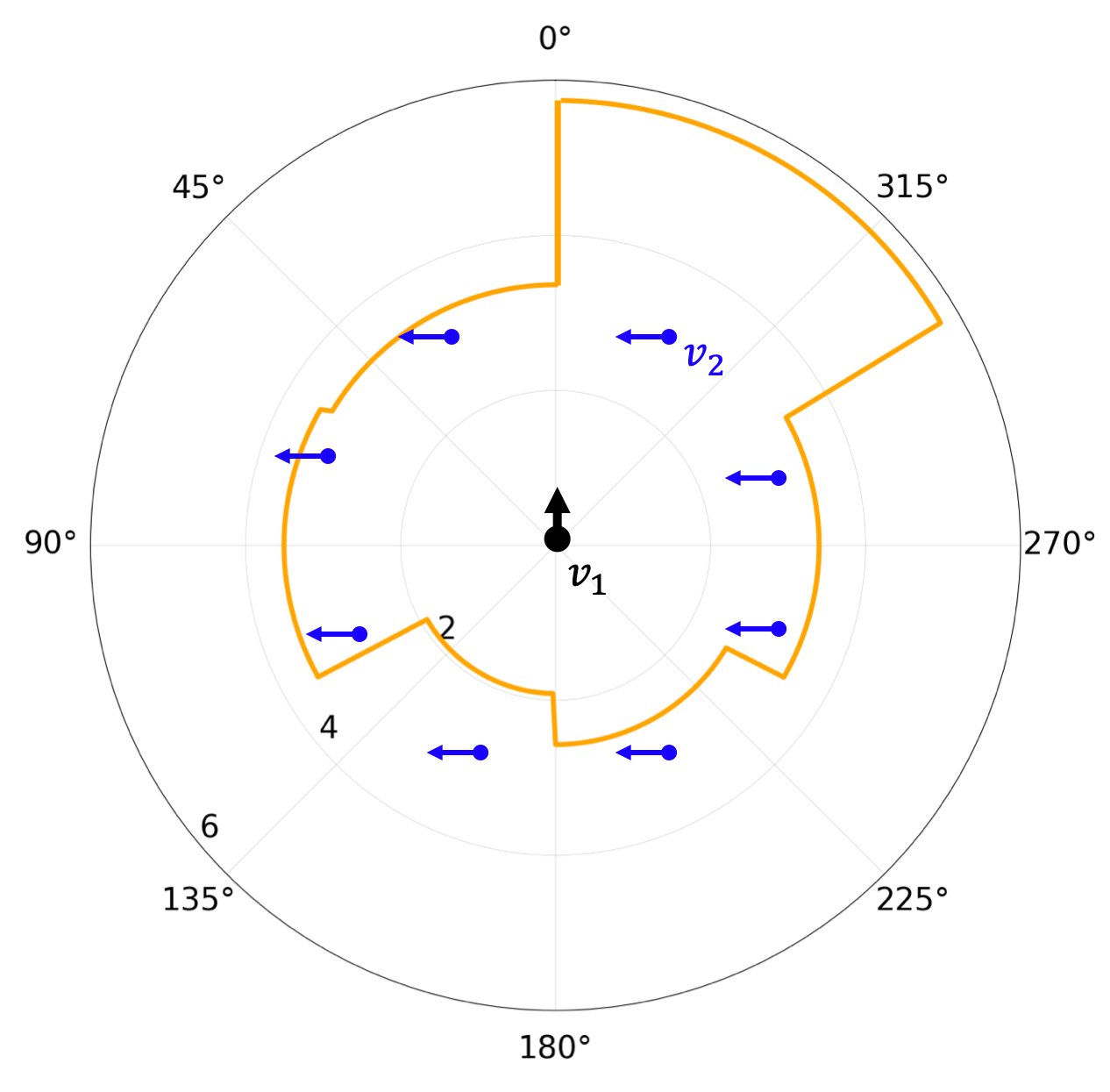}
        \caption{Learned Domain Parameter for $\phi_{21}$=90$^{o}$}
        \label{domain_90}
    \end{figure}

Prior literature on data-driven modeling intent of interacting agents, especially in the pedestrian domain, make strong assumptions such as all agents in a certain grid space influence each other's trajectories identically. However, pedestrians walking in crowded environments do not realistically pay equal attention to all agents at equal distances from the self. For instance, a pedestrian walking in a certain direction is most influenced by pedestrians walking in the same direction directly ahead of it, or towards it. Pedestrians are almost negligibly attentive towards pedestrians walking away from them or walking directly behind them. This behavior is also expected in other autonomous agents in crowded environments. By virtue of introducing a learnable vessel domain parameter, our model is able to differentiate and variably attend to different agents at the same distance from an agent, based on their relative headings and relative bearings from the self. The vessel domain parameter as learned by our spatially and temporally attentive model is shown in Figure \ref{domain_img}. In general, the model learns a farther distance from the self for relative bearings that fall in the line-of-sight of the vessel, and closer distances from the self for relative bearings that fall behind the vessel. Further, the model learns a farther distance for all neighbors $v_2$ that are approaching $v_1$ head-on, with $120^{o}\leq \phi_{21}<180^{o}$. This implies that between two neighbors, both at equal distances from $v_1$ and heading in the same direction, $v_1$ would be more influenced by the one that is approaching it head-on than another with the same relative heading but at a different relative bearing from $v_1$. Similarly, the model learns a farther distance for all neighbors $v_2$ with 60$^{o}\leq \phi_{21}$ $<$ 120$^{o}$ trying to cross $v_1$ from its starboard side, or present directly in front, as compared to other neighbors oriented in the same direction but at different relative bearing angles from $v_1$. Figure \ref{domain_90} shows the vessel domain as learned by the model for a vessel $v_2$ with $\phi_{21}$ = 90$^{o}$ for various $\theta_{21}$ values. As can be seen from the figure, the model attends more to $v_2$ when it tries to cross it from its starboard side, as compared to other relative bearings. This is understandable because neighbors with the same orientation at other relative bearings have no influence on its intent or high-level trajectory, and pose no immediate risk of collision to $v_1$. 
   \begin{figure}
       \centering
       \includegraphics[width=\textwidth]{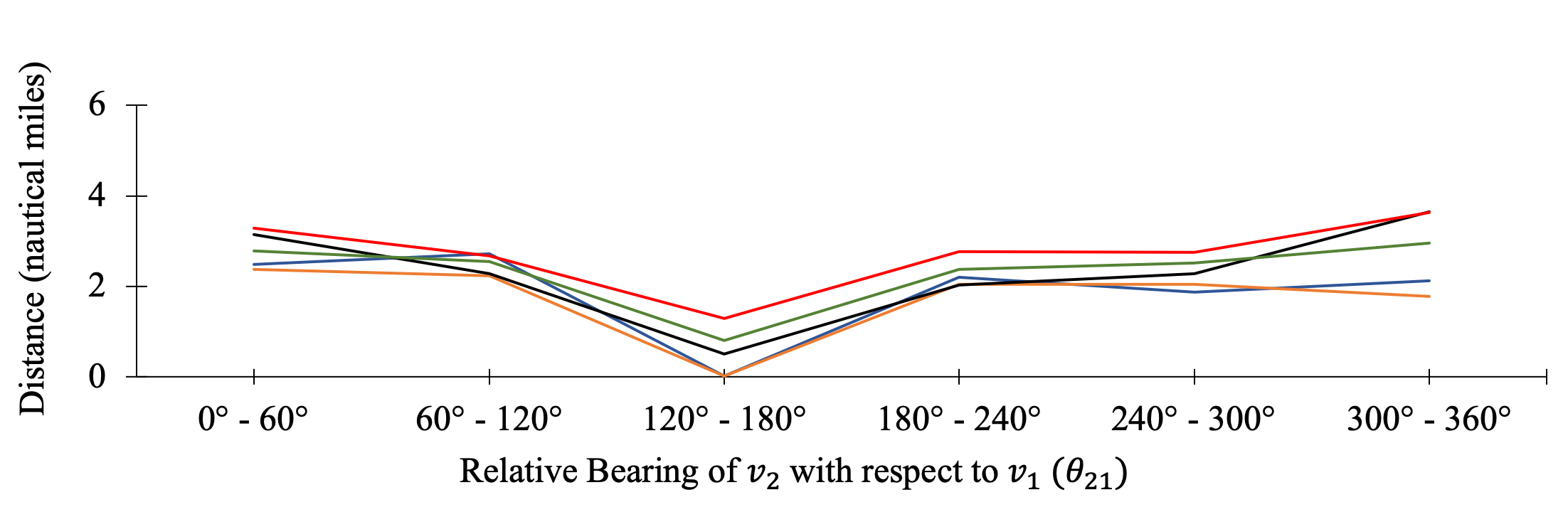}
       \caption{Robustness to Random Initializations 120$^{o}$ $<$ $\phi_{21}\leq 180^{o}$}
       \label{random_init}
   \end{figure}

By practice, deep neural networks are initialized to random weights before beginning the training process. Since this randomness causes the optimal parameter search to initiate at a different point and progress differently each time the model is trained on the same dataset, it may cause the model to converge at a different parameter configuration each time. Since we explicitly specify the logic of our spatial attention mechanism, we expect the model to understand and converge to similar values for the learn-able domain parameter, $S$, described in Section \ref{hw_attn} even for different training runs with different random initializations. Therefore, to evaluate the robustness of our model to randomness in learning and the ability to reproduce this domain parameter despite randomness, we train our model using 5 different random initialization seeds. Our model is able to learn a similar parameter across all five training runs. Figure \ref{random_init} shows the learned domain values for a scenario with a neighboring vessel $v_2$ at a relative heading  (120$^{o}$ $<$ $\phi_{21}\leq 180^{o}$) with respect to $v_1$ for 5 different random initializations. As can be seen from the figure, the model learns a nearly consistent value for the domain parameter across all the initializations. The learned value for $S$ is highest when the neighboring vessel $v_2$ with the given relative heading $\phi_{21}$ approaches $v_1$ head-on, i.e., -60$^{o}$ $<$ $\theta_{21}\leq 60^{o}$. The learned value for $S$ is lowest when $v_2$ is behind $v_1$ and heading away from $v_1$ with the given relative heading $\phi_{21}$.

\section{Conclusion}
In this work, we propose a learning-based method for modeling intent of vessels, hence enabling safe navigation in cluttered environments such as harbors. Despite being trained on only positional data, our novel architecture is able to accurately model vessel intent and is also able to infer knowledge such as vessel domain from observed data. Our model can be used alongwith other sophisticated data sources, such as sensors like LiDARs, radars, etc. for improved accuracy and user trust in safety-critical scenarios. While we validate our approach on the maritime domain, this method can be easily adopted to model intent and spatial interactions for other socially interacting autonomous agents, such as pedestrians, automobiles and unmanned aerial vehicles.

\bibliography{vessel.bib}

\end{document}